\documentstyle[twoside,fleqn,espcrc2,epsf,rotate]{article}
 
\newcommand{\vsbf}{\vspace*{-10pt}}
\newcommand{\vsaf}{\vspace*{-10pt}}
\newcommand{\vsbfc}{\vspace*{-50pt}}

\newcommand{\FIGUREFLOAT}[3]{%
  \begin{figure}[htbp]
    \vsbf
    \begin{center}
      \epsfxsize=\hsize
      \epsffile{#1}
    \end{center}
    \vsbfc
    \caption{\protect #3}
    \label{Fig:#2}
    \vsaf
  \end{figure}
}

\newcommand{\FIGURE}[3]{%
  \begin{figure}[htb]
    \vsbf
    \begin{center}
      \epsfxsize=\hsize
      \epsffile{#1}
    \end{center}
    \vsbfc
    \caption{\protect #3}
    \label{Fig:#2}
    \vsaf
  \end{figure}
}

\newcommand{\FIGURERIGHT}[3]{%
  \begin{figure}[htb]
    \vsbf
    \begin{center}
      \epsfysize=\hsize
      \rotate[r]{\epsfbox{#1}}
    \end{center}
    \vsbfc
    \caption{\protect #3}
    \label{Fig:#2}
    \vsaf
  \end{figure}
}

\begin{document}
 
\title{%
Status of Heavy Quark Physics on the Lattice
}
 
\author{Terrence Draper 
  \address{%
    Department of Physics and Astronomy,
    University of Kentucky, Lexington, KY 40506, USA
  }
}
 
\begin{abstract}
The status of lattice calculations of some phenomenology of heavy quarks
is presented.  Emphasis is on progress made in calculating those
quantities relevant to estimating parameters of the quark mixing matrix,
namely leptonic decay constants, the bag parameter of neutral $B$
mixing, and semileptonic form factors.  New results from studies of
quarkonia are highlighted.
\end{abstract}
 
\maketitle

\section{INTRODUCTION}

Lattice QCD offers the best hope for estimating the non-perturbative QCD
effects in weak decays of hadrons, thereby allowing the determination of
elements of the Cabibbo-Kobayashi-Maskawa (CKM) matrix from experimental
data.  Much of the phenomenological interest is in those weak decays
which contain at least one heavy quark, but it is this case where
technical issues are often most difficult and conceptual points most
subtle.

There are a variety of approaches taken to simulate a heavy quark on
today's lattices.  One is to use the improved actions used for
light-quark calculations, taking care to keep $am$ not too large, and
then to rely on an extrapolation (often guided by the result of a static
calculation) in the heavy quark mass from the charm region to the
bottom.  A philosophically-different approach is that of
non-relativistic QCD (NRQCD), namely, to integrate out the heavy quark
producing an effective action which is then discretized; then $am$ must
be kept large.  A third approach, developed at Fermilab, seeks to
produce an action which can be used for {\it any\/} value of $am$, with
systematic errors that can be identified and controlled.  An
approximation to the full program is often used wherein standard
relativistic actions are reinterpreted non-relativistically.  Much of
the debate focuses on to what extent systematic errors are controlled in
this case.  J. Sloan gave a nice compare-and-contrast summary of NRQCD
and Fermilab-type actions at {\sl Lattice '94}~\cite{Slo_lat94}.

I review developments made and results announced this past year.  Sec. 2
summarizes recent calculations of the leptonic decay constants, notably
$f_B$, where there has been considerable progress.  In Sec. 3, I present
new results, and new analysis of old data, for the $B_B$ parameter of
neutral $B$ mixing.  Sec. 4 outlines the successes and limitations of
calculations of semileptonic form factors for $B$ and $D$ decays; this
year has seen the initiation of some new large-scale projects.  Sec. 5
highlights new developments and results of calculations, presented at
this conference, of the phenomenology of quarkonia.  I conclude with a
qualitative overview of where we stand, and in what direction future
work may be focused.

\section{LEPTONIC DECAY CONSTANTS}

It has been a banner year for the calculation of $f_{B}$, the leptonic
decay constant of the $B$-meson.  Several large groups, having invested
many person-years of effort, have presented final results this year;
many of these calculations have done a comprehensive job of estimating
systematic errors.  There appears to be a consensus among many of the
groups (See Figure~\ref{Fig:fb}.)  This is encouraging, given the
diversity of actions and analyses. See reference~\cite{Ono_lat97} for a
discussion of discretization errors.
\FIGURE
  {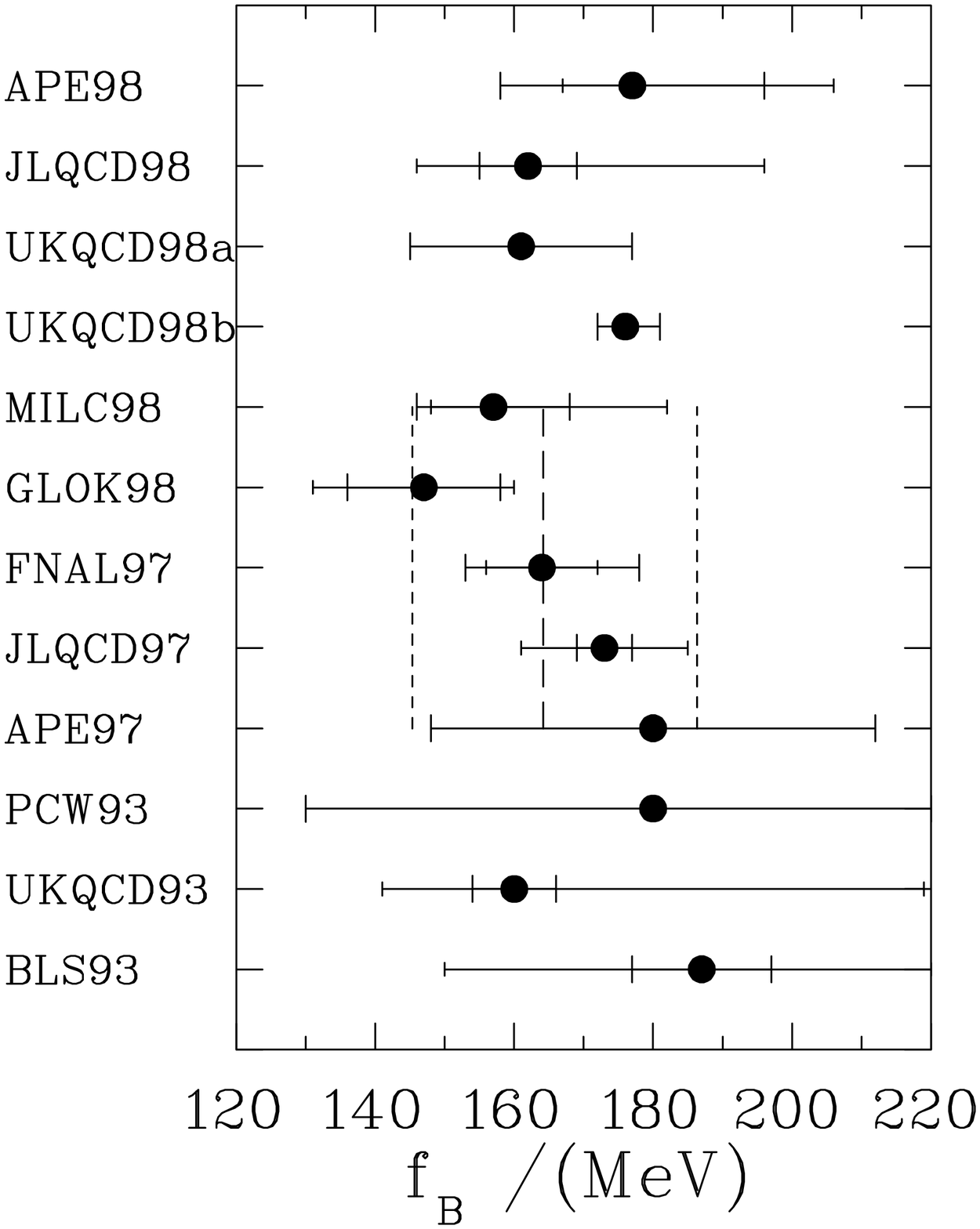}
  {fb}
  {A summary of recent lattice results for $f_{B}$.  Data and labels are
   taken from Table~1.  The error bars with the longer feet are
   statistical.  Systematic errors have been combined in quadrature.
   Also shown is a world average over the subset of results which have
   final and complete estimates of systematic errors.}

The MILC collaboration has published their final results this
year~\cite{MILC_hp9806412}. They use Wilson quarks for both the light
and heavy quarks, but use some aspects of the ``Fermilab
interpretation''~\cite{FNAL_hp9711426}, namely the use of
non-relativistic normalization and kinetic, rather than pole mass, for
the heavy quark.  MILC has also found it useful to include static-Wilson
results to aid in the extrapolation in the heavy quark mass.  With
Wilson heavy quarks in the Fermilab interpretation, there is a ${\cal
O}(1/M)$ error (physical, {\it not\/} a lattice artifact) which cannot
be extrapolated away.  MILC, following JLQCD~\cite{JLQCD_hl9711041},
estimates this effect to be $\approx 2\%$ from a tree-level
approximation.  MILC has been very careful in estimating systematic
errors; changes since their preliminary {\sl Lattice '97\/} results
include updates of their chiral extrapolation, the method of combining
errors within the quenched approximation, and some new unquenched data
to estimate quenching errors.

In T. Onogi's {\sl Lattice '97\/} review~\cite{Ono_lat97}, he reported
that although the MILC and JLQCD raw data agreed for common parameters,
there were some residual differences from the way the scale was set from
$f_{\pi}$.  These discrepancies were discussed after last year's
symposium by members of each group, and the discrepancies have now been
resolved~\cite{Ber-pc}.

The Fermilab (FNAL) collaboration have published their final
results~\cite{FNAL_hp9711426}, updated since {\sl Lattice '97}.  They
used the SW (clover action) for both heavy and light quarks, but
interpreted non-relativistically; this includes the use of correctly
normalized fields, an additional three-dimensional rotation for the
heavy quark in the current, and the use of the kinetic mass to set the
quark mass.  They include the full ${\cal O}(a)$ correction, but not the
complete ${\cal O}(\alpha a)$ correction to the current.  The advantage
of their program is that they can ``sit on'' the $b$ mass. They state
that their systematic errors are under control, are smaller than their
systematic errors, and are smaller than the quenching error which they
estimate to be about $10\%$.

The GLOK Collaboration has published their final
results~\cite{GLOK_hl9801038} updated since {\sl Lattice '97\/} (see
also~\cite{Bha_lat98}).  Their's is a quenched, $\beta=6.0$, calculation
which uses clover light (tree-level tadpole improved) and NRQCD heavy
quarks.  The Hamiltonian includes all ${\cal O}(1/M^2)$ corrections, the
leading ${\cal O}(1/M^3)$ correction, and discretization corrections at
the same level.  One important ingredient in the calculation was their
axial current renormalization~\cite{MorShi_hl9712016}.  This was the
first use of mass-dependent matching factors and ${\cal O}(\alpha a)$
current correction. They found mixing between lattice currents which
does not vanish as $M\rightarrow\infty$ because of an ${\cal O}(\alpha
a)$ lattice artifact term, which can be absorbed into an ${\cal
O}(\alpha a\Lambda_{QCD})$ discretization correction.  The latter is
anomalously large (they found $\approx 12\%$ reduction at the $B$ mass).
Furthermore, it cannot be included consistently if Wilson light quarks
are used, that is, if ${\cal O}(a)$ corrections are included for heavy
quarks, they must be included for the light quarks as well --- one is
thus left with sizeable scaling violations if one uses NRQCD-heavy with
Wilson-light quarks.  GLOK has augmented their $\beta=6.0$ calculation
with a $\beta=5.7$ run in order to assess scaling
violations~\cite{Hei_lat98}.

Figure~\ref{Fig:FNAL_GLOK_compare} compares these NRQCD results
of~\cite{GLOK_hl9801038} with those of Fermilab~\cite{FNAL_hp9711426}.
For the $B$ meson system, the actions are very similar.  The results
ought to agree and they do.
\FIGURE
  {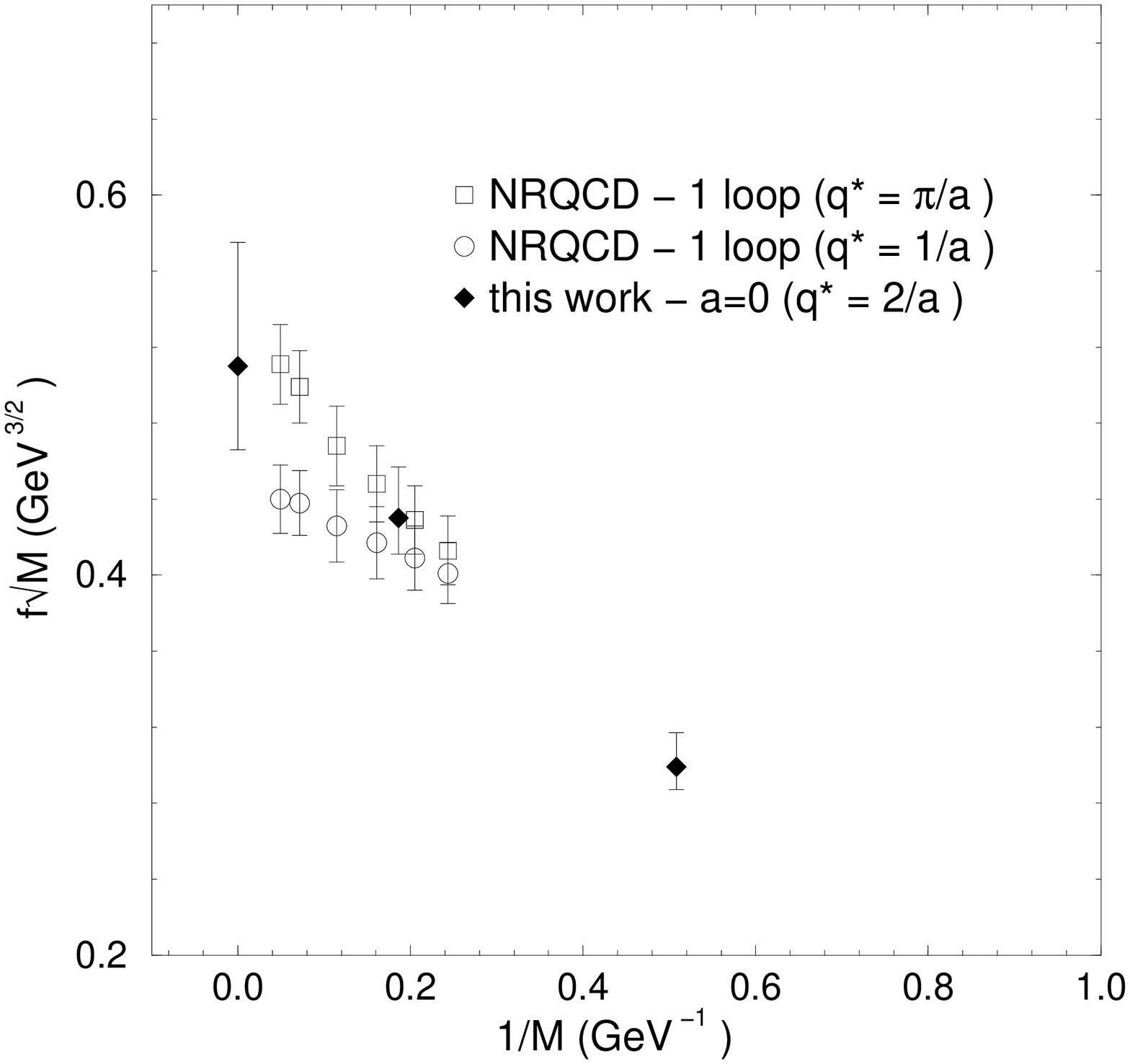}
  {FNAL_GLOK_compare}
  {A comparison by Fermilab~\cite{FNAL_hp9711426} of their results with the
   NRQCD results of GLOK~\cite{GLOK_hl9801038}.}

JLQCD~\cite{JLQCD_lat98} has presented preliminary results at this
conference, using actions and analysis which follow GLOK's lead.  They
use quenched Wilson gauge actions at $\beta=5.7$, $5.9$ and $6.1$, SW
light quarks with a clover coefficient calculated at one-loop, and NRQCD
heavy quarks, including both ${\cal O}(1/M)$ and ${\cal O}(1/M^2)$
terms.  Their results confirm the features seen by GLOK; namely, the
contribution from operator mixing of the temporal component of the axial
vector current is larger than one would expect~\cite{MorShi_hl9712016},
and this softens the $a$-dependence of $f_{B}$, the $q^*$ dependence is
large at one-loop, but is expected to be small at two-loop, and the
one-loop correction softens the slope of $f_{B}$ versus $1/M_{B}$.
\FIGURERIGHT
  {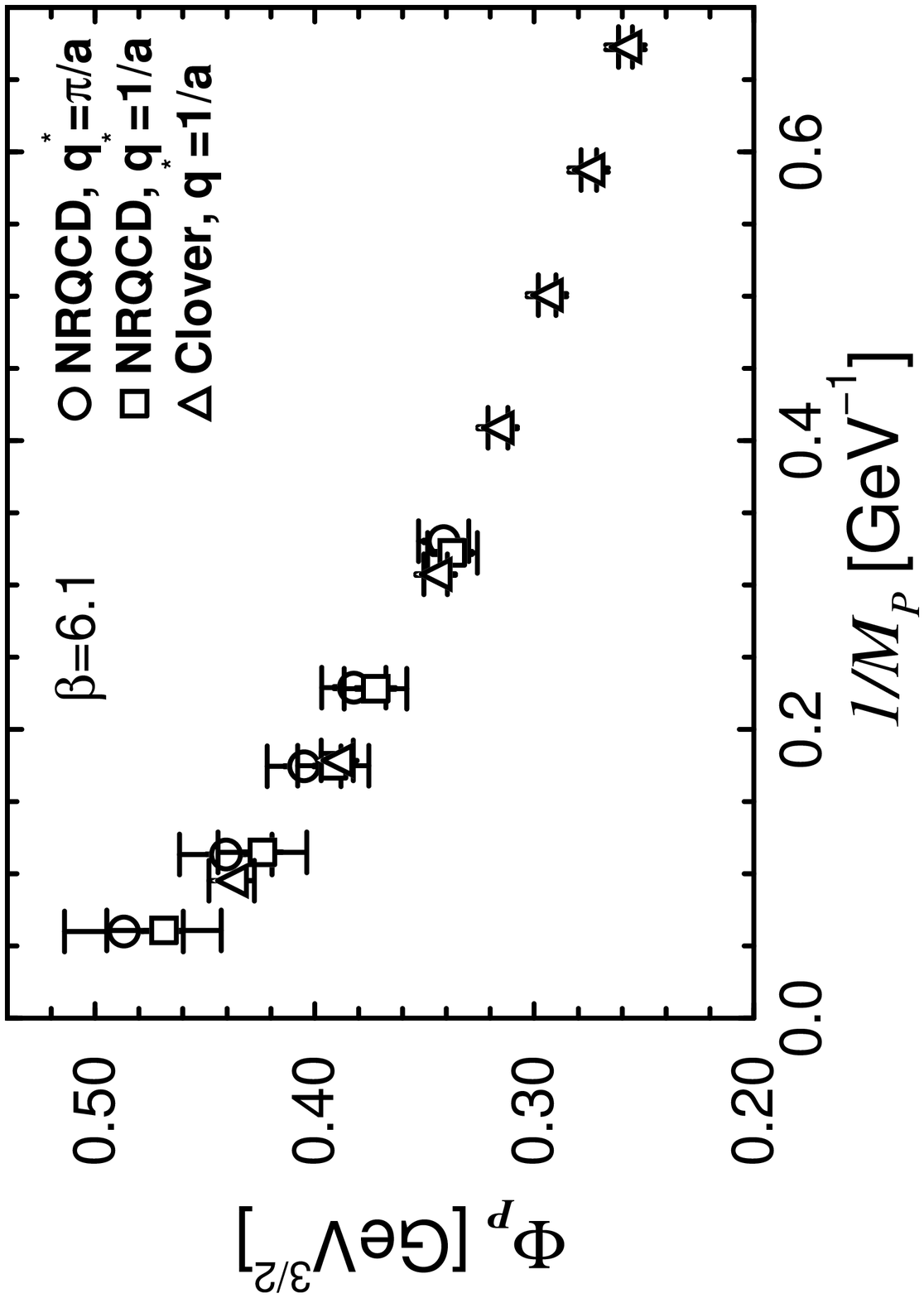}
  {JLQCD_ClvsNR}
  {$1/M_B$ dependence of $\Phi_B$ from NRQCD (for two choices of $q^*$) and
   Heavy-Clover at $\beta=6.1$.
   From~\cite{JLQCD_lat98}.}
Figure~\ref{Fig:JLQCD_ClvsNR} shows their comparison of these
preliminary clover-light--NRQCD-heavy data with their
clover-light--clover-heavy data presented last
year~\cite{JLQCD_hl9711041}. They are consistent.

At this conference, Lin and Lellouch~\cite{Lin_lat98} presented preliminary
high-statistics results of leptonic decay constants using a tadpole-improved
tree-level SW quark action for both heavy and light quarks (with KLM
normalization for the quark fields).  They ran at $\beta=6.0$ on a $16^3\times
48$ lattice and at $\beta=6.2$ on $24^3\times 48$.  They do not attempt a
continuum extrapolation, so I, as they, will take their $\beta=6.2$ data as
their best.

The ALPHA collaboration~\cite{ALPHA_hl9605038} has proposed a method for
systematic improvement of the Wilson action and bilinear quark operators
to remove all ${\cal O}(a)$ discretization errors.  The program
evaluates the clover coefficient and renormalization constants (of
bilinear quark operators) non-perturbatively.  This year, two new
projects have calculated heavy-light leptonic decay constants using this
approach.  APE~\cite{Bec_lat98} and UKQCD~\cite{Ric_lat98} ran at
$\beta=6.2$ on $24^{3}\times 64$ and $24^{3}\times 48$ lattices,
respectively.  Both simulate with the heavy quark near charm to keep
${\cal O}(am)$ errors small; thus they obtain $f_{D}$ by interpolation.
However, APE finds that the extrapolation up to the bottom region is
quite sensitive to the functional form of the fit; this inflates their
systematic error.  In these proceedings, UKQCD quotes only statistical
errors.

\begin{table*}[t]
\begin{tabular}{llll}
\hline                            & $f_{B}/{\rm MeV}$               &  $f_{B_{s}}/{\rm MeV}$               & $f_{B_{s}}/f_{B}$                    \\
\hline                   
APE98~\cite{Bec_lat98}            & $177(19)(^{29}_{10})$           & $206(16)(^{32}_{0})$                 & $1.16(4)$                            \\   
JLQCD98~\cite{JLQCD_lat98}        & $162(^{7}_{7})(^{34}_{16})$     & $190(^{5}_{5})(^{42}_{18})$          & $1.18(3)(^{5}_{5})$                  \\
UKQCD98a~\cite{Lin_lat98}         & $161(16)$                       & $190(12)$                            & $1.18(8)$                            \\
UKQCD98b~\cite{Ric_lat98}         & $176(^{5}_{4})$                 &                                      &                                      \\
\hline
MILC98~\cite{MILC_hp9806412}      & $157(11)(^{25}_{9})(^{23}_{0})$ & $171(10)(^{34}_{9})(^{27}_{2})$     & $1.11(^{2}_{2})(^{4}_{3})(^{3}_{3})$ \\
GLOK98~\cite{GLOK_hl9801038}      & $147(11)(^{8}_{12})(9)(6)$      & $175(8)(^{7}_{10})(11)(7)(^{7}_{0})$ & $1.20(4)(^{4}_{0})$                  \\
FNAL97~\cite{FNAL_hp9711426}      & $164(^{14}_{11})(8)$            & $185(^{13}_{8})(9)$                  & $1.13(^{5}_{4})$                     \\
JLQCD97~\cite{JLQCD_hl9711041}    & $173(4)(12)$                    & $199(3)(14)$                         &                                      \\
APE97~\cite{APE_hl970300}         & $180(32)$                       & $205(35)$                            & $1.14(8)$                            \\
\hline
PCW93~\cite{PCW_hl9312051}        & $180(50)$                       &                                      & $1.09(2)(5)$                         \\
UKQCD93~\cite{UKQCD93}            & $160(^{6}_{6})(^{59}_{19})$     & $194(^{6}_{5})(^{62}_{9})$           & $1.22(^{4}_{3})$                     \\
BLS93~\cite{BerLabSon_hl9306009}  & $187(10)(34)(15)$               & $207(9)(34)(22)$                     & $1.11(6)$                            \\
\hline
\end{tabular}
\label{Table:fb}
\caption{Results for $f_{B}$, $f_{B_{s}}$ and $f_{B_{s}}/f_{B}$ obtained in the
quenched approximation.  The errors are statistical, then various
systematic.  Results are divided into three sections.  In the top
section are the most recent {\it preliminary\/} results; thus, despite
their quality they are not included in world averages.  In the middle
section are recent published and {\it finalized\/} results where all
systematic errors (except quenching) have been carefully assessed; these
are folded into a world average.  In the last section are selected older
results from pre-modern methods.}
\end{table*}

\begin{table*}[t]
\begin{tabular}{llll}
\hline
                                  & $f_{D}/{\rm MeV}$               &  $f_{D_{s}}/{\rm MeV}$               & $f_{D_{s}}/f_{D}$                    \\
\hline
APE98~\cite{Bec_lat98}            & $202(14)(^{0}_{12})$            &  $231(11)(^{7}_{0})$                 & $1.11(3)$                            \\   
UKQCD98a~\cite{Lin_lat98}         & $193(10)$                       &  $221(9)$                            & $1.15(4)$                            \\   
UKQCD98b~\cite{Ric_lat98}         & $190(^{5}_{2})$                 &                                      &                                      \\   
\hline
MILC98~\cite{MILC_hp9806412}      & $192(11)(^{16}_{8})(^{15}_{0})$ &  $210(9)(^{25}_{9})(^{17}_{1})$      & $1.10(^{2}_{2})(^{4}_{2})(^{2}_{3})$ \\   
FNAL97~\cite{FNAL_hp9711426}      & $194(^{14}_{10})(10)$           &  $213(^{14}_{11})(11)$               & $1.10(^{4}_{3})$                    \\   
JLQCD97~\cite{JLQCD_hl9711041}    & $197(2)(17)$                    &  $224(2)(19)$                        &                                      \\   
APE97~\cite{APE_hl970300}         & $221(17)$                       &  $237(16)$                           & $1.07(4)$                            \\   
LANL95~\cite{BhaGup_hl9510044}    & $186(29)$                       &  $218(15)$                           &                                      \\   
\hline
PCW93~\cite{PCW_hl9312051}        & $170(30)$                       &                                      & $1.09(2)(5)$                         \\   
UKQCD93~\cite{UKQCD93}            & $185(^{4}_{3})(^{42}_{7})$      &  $212(^{4}_{4})(^{46}_{7})$          & $1.18(^{2}_{2})$                     \\   
BLS93~\cite{BerLabSon_hl9306009}  & $208(9)(35)(12)$                &  $230(7)(30)(18)$                    & $1.11(6)$                            \\
\hline
\end{tabular}
\label{Table:fd}
\caption{Same as for Table~1 but for $f_{D}$, $f_{D_{s}}$, $f_{D_{s}}/f_{D}$.}
\end{table*}
   
Tables~1 and~2 list and Figure~\ref{Fig:fb} displays selected results
for heavy-light decay constants and ratios.  In computing a world
average (Table~3), always a notoriously difficult thing to attempt, I
include only those final, not preliminary, large-scale results where
modern methods have been used and a comprehensive investigation of
systematic errors has been done, which includes either a continuum
extrapolation, or an estimate of scaling errors.  Also tabulated and
graphed are a sampling of other results.  Emphasis is on recent results,
especially those preliminary results presented at this conference.  See
reference~\cite{FlySac_hl9710057} for a more comprehensive listing of
older results.
\begin{table}
\begin{center}
\begin{tabular}{lll}
\hline
                        & Quenched          & ``Unquenched''     \\
\hline					    
\\					    
$f_{B}/\,{\rm MeV}$     & $165^{+20}_{-20}$ & $185^{+35}_{-25}$  \\
$f_{B_s}/\,{\rm MeV}$   & $185^{+25}_{-20}$ & $210^{+40}_{-25}$  \\
$f_{D}/\,{\rm MeV}$     & $200^{+20}_{-20}$ & $215^{+30}_{-25}$  \\
$f_{D_s}/\,{\rm MeV}$   & $220^{+25}_{-20}$ & $240^{+30}_{-25}$  \\
$f_{B_s}/f_{B}$         & $1.14^{+6}_{-5}$  & $1.14^{+7}_{-6}$   \\
$f_{D_s}/f_{D}$         & $1.10^{+6}_{-4}$  & $1.11^{+6}_{-5}$   \\
\hline
\end{tabular}
\end{center}
\label{Table:fb_average}
\caption{World averages for quenched $f_{B}$, $f_{B_{s}}$, $f_{D}$,
$f_{D_{s}}$, and ratios.  The ``unquenched'' results are best {\it
guesses\/} obtained by shifting the quenched values by the rough
estimates of quenching errors from MILC~\cite{MILC_hp9806412} (the last
of their errors in Tables~1 and~2).  Estimates of unquenched errors
assume that the MILC corrections could be off by a factor of 2.}
\end{table}

Also reported at this conference were two investigations into the
spectrum of heavy-light mesons.  Ali Khan~\cite{Ali_lat98} reported on
results from an extensive calculation of $B$, $B_c$ $\Upsilon$, and
$b$-baryon spectra; the study included radially and orbitally excited
mesons.  Lewis and Woloshyn~\cite{LewWol_hl9803004} studied the relative
effect of $1/M$, $1/M^2$ and $1/M^3$ terms from the NRQCD heavy
quark. The expansion is well behaved for bottom; effects are larger, as
expected, for charm.

In summary, leptonic decay constant calculations are quite mature;
several large groups and years of effort have resulted in results with
small statistical errors, and systematic errors which, aside from
quenching, are largely identified and controlled.  Pleasingly, there
appears to be a consensus on values among those groups that have
recently published their final results.  (This consensus needs to be
confirmed by the new projects which use non-perturbative estimates of
coefficients and extrapolate up from lighter masses.)  The world average
of $f_B$ in the quenched approximation, obtained from from various
actions and groups, has stabilized in last 2--3 years (world averages
had been
reported~\cite{Ono_lat97,Ber_hp9709460,FlySac_hl9710057,Wit_hl9705034,Fly_lat96}
in the range $160$--$175\,{\rm MeV}$ with errors of about
$25$--$35\,{\rm MeV}$), and is significantly lower than several years
ago (when world averages were $200(40)\,{\rm MeV}$~\cite{All_lat95} or
above).  Looking back~\cite{FNAL_hp9711426}, one sees that older results
were higher as several effects conspired: old static values which guided
extrapolations in $1/M_{B}$ were misleadingly high prior to modern
variational techniques which remove excited-state contaminations, and
both $1/M_B$ corrections to the static limit and discretization errors
are larger than were expected.  Quenching errors currently are difficult
to estimate.  Unquenching may increase $f_B$ by $10$--$15\%$ (MILC),
maybe more~\cite{SGO98}, or maybe less~\cite{Sha_lat98}.  Lattice
estimates of $f_{B}$ continue to be very useful phenomenologically.  In
the future, we'll see continued work with improved actions and
renormalization, with the emphasis shifting towards making better
estimates of quenching errors.

\section{NEUTRAL $B$ MIXING}

The CKM matrix element $|V_{td}|$ can be extracted from the mass
difference induced in neutral $B$ mixing
\begin{eqnarray*}
(\Delta m)|_{B-\overline{B}} = f_{B}^{2} B_{B} |V_{td}|^{2} F(m_{t},M_{W})
\end{eqnarray*}
where $F$ is a known function and the $B_{B}$ parameter has been used to
parameterize the matrix element $\langle\overline{B}^{0} | {\cal O}_{L} | B^{0}
\rangle$ of the $\Delta B=2$ four-quark operator ${\cal O}_{L} =
\overline{b}\gamma_{\mu}(1-\gamma_{5})q \,
\overline{b}\gamma_{\mu}(1-\gamma_5)q$ in terms of its approximation under the
factorization hypothesis (``vacuum saturation'' or VSA)
\begin{eqnarray*}
      B_{B}
 & = &   
      \frac{
\langle\overline{B}^{0} | {\cal O}_{L} | B^{0} \rangle _{\phantom{\rm VSA}}
}
{
\langle\overline{B}^{0} | {\cal O}_{L} | B^{0} \rangle _{\rm VSA}
} 
 = 
      \frac{
\langle\overline{B}^{0} | {\cal O}_{L} | B^{0} \rangle 
}
{
8/3 f_{B}^{2}M_{B}^{2}
}
\label{B_B_def}
\end{eqnarray*}

Since one needs a non-perturbative evaluation of the matrix element and
thus the combination $f_{B}^{2} B_{B}$, why calculate $f_{B}$ and $B_B$
separately?  Reasons include (1) a quite precise value can be obtained,
with an optimal choice of smearing function, from relatively few
configurations, because $B_{B}^{}$ can be extracted from a ratio of
three- to two-point functions which are strongly correlated, (2)
perturbative corrections tend to be stabilized because of cancelations
in numerator and denominator, and (3) it seems as though VSA may be a
surprisingly good approximation (to $\approx 10$--$15\%$) for the
$B_{B}^{}$ parameter; this is an important qualitative statement, of use
to model builders, which should not be obscured by poor-statistics
attempts to calculate the product $B_{B}^{} f_{B}^{2}$.

The $B_B$ parameter was reviewed by J. Flynn~\cite{Fly_lat96} at {\sl
Lattice '96\/} and by A. Soni~\cite{Son_lat95} at {\sl Lattice '95}.
Results from UKQCD (Lin and Lellouch)~\cite{Lin_lat98}, and from the
Hiroshima group~\cite{Yam_lat98}, were reported at this conference.
Other recent results include those from Di~Pierro and
Sachrajda~\cite{DiPSac_hl9805028}, from Gim\'{e}nez and
Reyes~\cite{GimRey_hl9806023}, and from Bernard, Blum and
Soni~\cite{BerBluSon_hl9801039}.

This year, Lin and Lellouch~\cite{Lin_lat98} have calculated the $B_B$
parameter as part of larger effort to calculate $B$ parameters and decay
constants for light and heavy mesons.  They use the SW action, with the
clover coefficient tadpole improved at tree-level, at both $\beta=6.0$
and $6.2$ to check scaling.  They see little $a$-dependence.  In fact,
the world's collection of results from conventional (non-static) methods
(Figure~\ref{Fig:Bparam}) (the top set of results) shows consistency
among groups, no $a$-dependence and agreement between Wilson and SW
calculations.

\FIGURE
  {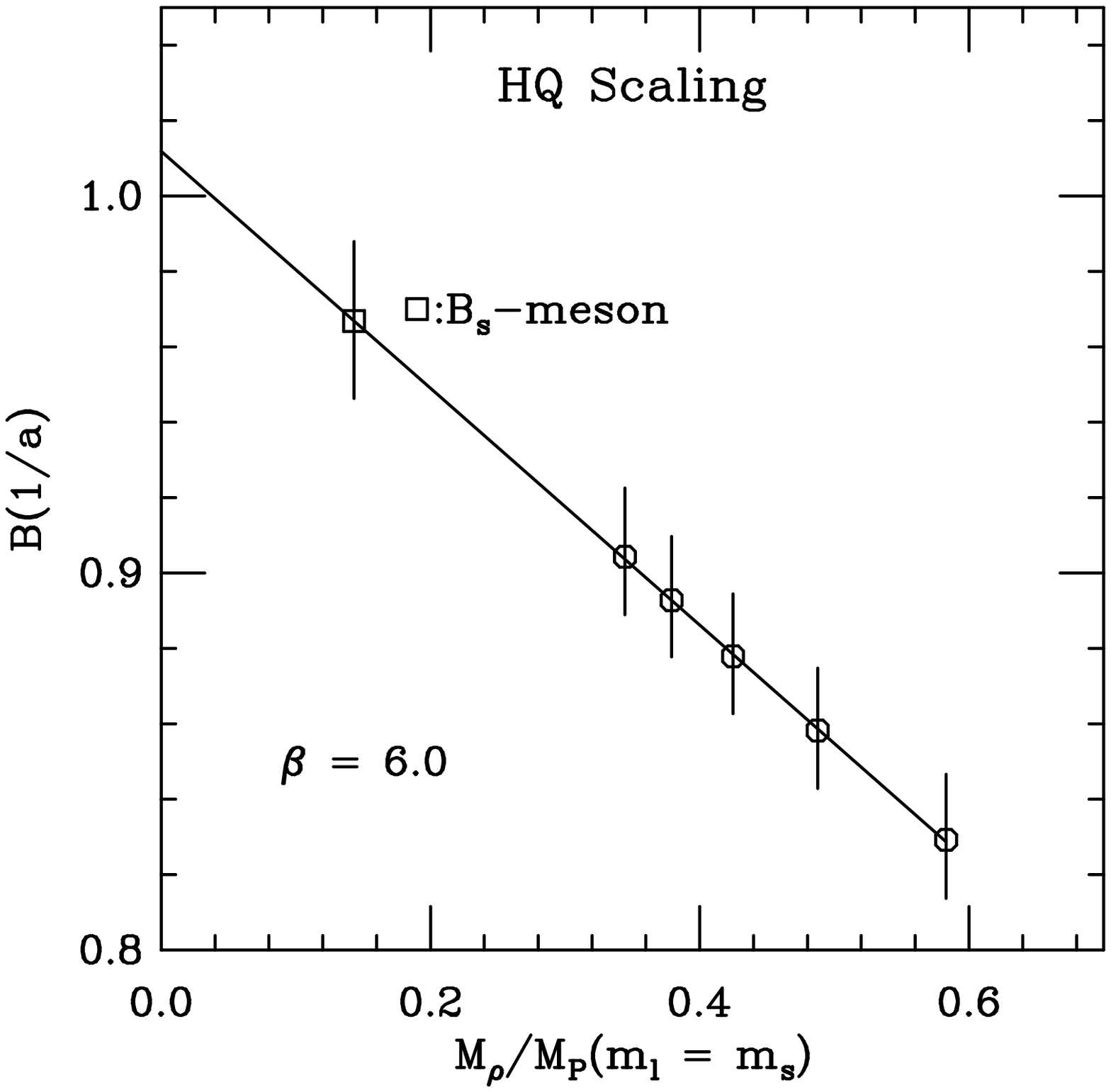}
  {Lin_Bparam}
  {Heavy quark mass dependence of $B_{B_s}$.
   From UKQCD (Lellouch/Lin)~\cite{Lin_lat98}.}
Figure~\ref{Fig:Lin_Bparam} shows the heavy quark mass dependence of
$B_{B_{s}}$.  There is a consensus~\cite{Lin_lat98,Son_lat95} that the
extrapolated-static value is larger than that at the physical mass.

\FIGUREFLOAT 
  {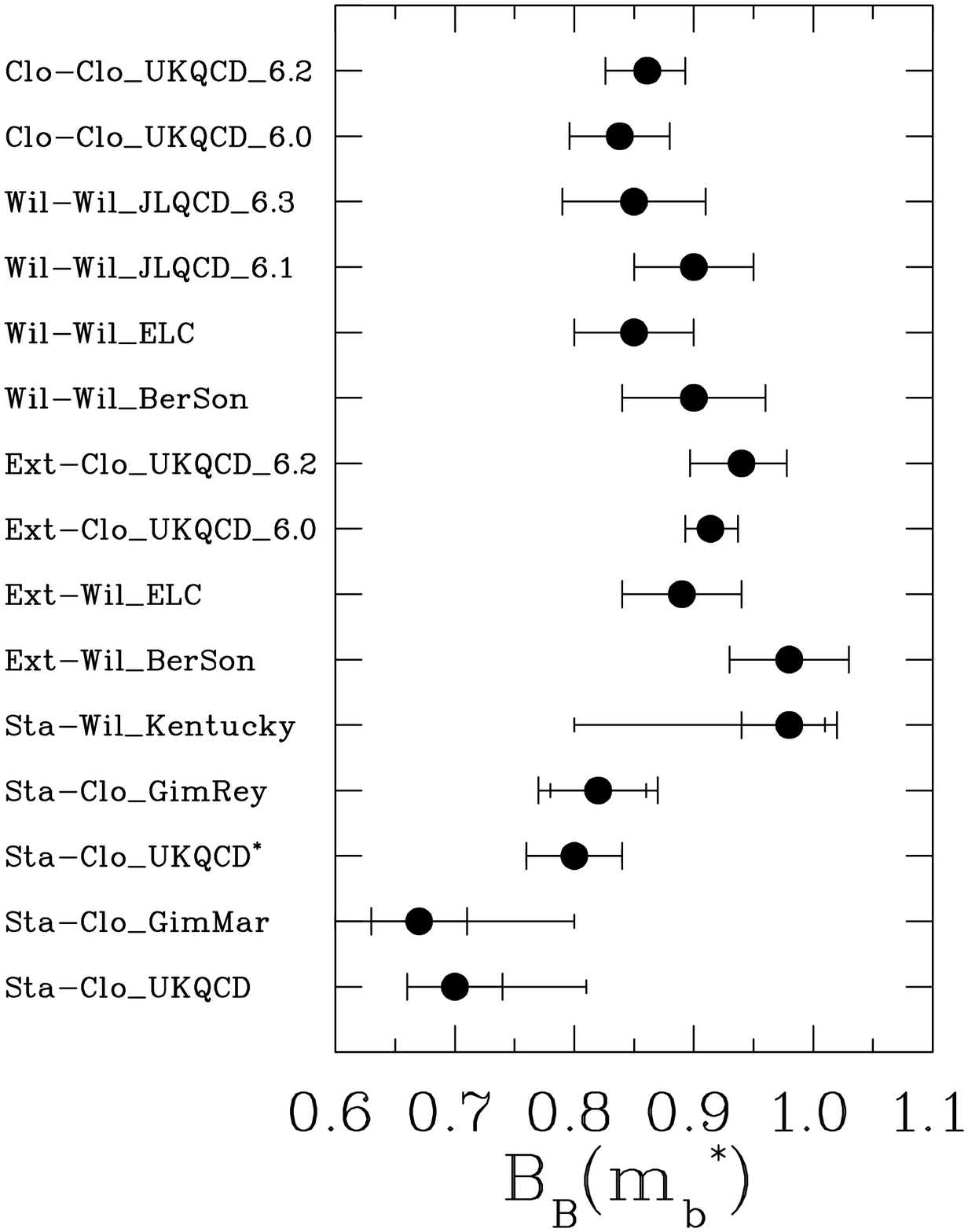} {Bparam} 
  {World results for the $B_{B}(m_b(m_b))$
  parameter.  For a more complete listing, see tables
  in~\cite{ChrDraMcN_hl9610026} and~\cite{FlySac_hl9710057,Fly_lat96}.
  ``Conventional methods'' by UKQCD~\cite{Lin_lat98},
  JLQCD~\cite{JLQCD_lat95}, ELC~\cite{ELC92}, and Bernard \&
  Soni~\cite{Son_lat95} labeled ``Clo-Clo'' or ``Wil-Wil'' use SW
  (clover) or Wilson, respectively, for both heavy and light quarks; the
  world average is taken from among this set.  ``Ext-Clo'' and
  ``Ext-Wil'' are the extrapolations of these data to the static limit
  for the heavy quark.  To these should be compared results of using
  directly the static approximation for the heavy quark (from
  Kentucky~\cite{ChrDraMcN_hl9610026}, UKQCD~\cite{UKQCD_hl9508030} and
  Gim\'{e}nez \& Martinelli~\cite{GimMar_hl9610024}) which are labeled
  ``Sta-Clo'' for a clover, or ``Sta-Wil'' for a Wilson, light quark.
  Gim\'{e}nez and Reyes~\cite{GimRey_hl9806023} have reanalyzed the
  renormalization of the old data of reference~\cite{GimMar_hl9610024}
  and of reference~\cite{UKQCD_hl9508030} (labeled UKQCD${}^*$).
  Di~Pierro and Sachrajda~\cite{DiPSac_hl9805028} have made a similar
  reanalysis and reach the same conclusions.}

The extrapolated-static results and those obtained directly in the
static approximation should agree.  There are disagreements among the
static calculations, some of which do not agree with extrapolated-static
results (Figure~\ref{Fig:Bparam}).  As with decay constant calculations,
one must fight the very poor signal-to-noise ratio inherent with using
static quarks.  A large number of configurations has to be generated if
one uses canonical smearing techniques~\cite{GimMar_hl9610024}.  The
Kentucky group obtains~\cite{ChrDraMcN_hl9610026} the same statistical
error with an order of magnitude fewer configurations by using optimal
smearing techniques that they developed~\cite{DraMcN_hl9401013}.  It is
notable that all groups' raw lattice data are consistent; furthermore,
the VSA holds surprisingly well for all operators (with appropriate
normalization).

As a short digression, it is interesting to note that factorization
seems to hold well in related static calculations.  Di~Pierro and
Sachrajda~\cite{DiPSac_hl9805028} need to evaluate matrix elements like
$\left\langle B\right| \ \overline{b}\gamma_{\mu} ( 1 - \gamma_5 ) q\
\overline{q}\gamma_{\mu}( 1 - \gamma_5 ) b\ \left| B\right\rangle$ for their
studies of inclusive decays of heavy hadrons.  The operators are similar to,
but different than, those for the $B_{B}$ parameter.  For the ``figure-8''
contractions, they find that factorization holds surprisingly well --- to
$\approx 5\%$.

For the static $B_B$ parameter calculations, differences in renormalized
quantities arise from choices made in lattice-to-continuum matching:
whether products and quotients are expanded in $\alpha$ or merely
multiplied and divided makes a uncomfortably large difference.

Although all organizations of perturbation theory at one-loop are
theoretically equal, some are more equal than others!  The Kentucky
group~\cite{ChrDraMcN_hl9610026} has been advocating an organization of
lattice perturbation theory which uses the
Lepage-Mackenzie~\cite{LepMac_hl9209022} choices for $\alpha_s$ and
tadpole improvement and which, by allowing explicit cancelations, is
insensitive to the wave-function normalization and which does not mask
the agreement of the raw data with VSA. (This also reduces the
statistical errors.)  Their results agree with VSA and with the
extrapolated static results and can thus be used in conjunction with
conventional data to interpolate to the $B$-meson mass, as has been done
with much success for the decay constant calculations.  The original
clover-light--static-heavy
results~\cite{UKQCD_hl9508030,GimMar_hl9610024} which disagreed with the
Kentucky results and with the world's extrapolated-static results, have
recently been reanalyzed~\cite{DiPSac_hl9805028,GimRey_hl9806023} with
an organization of perturbation theory which includes tadpole
improvement (and corrects errors in some renormalization constants); the
discrepancies have been reduced but not eliminated.

The Hiroshima group~\cite{Yam_lat98} has begun the first calculation of
the $B_{B}$ parameter using NRQCD for the heavy quarks (with $16^3\times
48$ quenched Wilson $\beta=5.9$ gauge-field configurations, and SW light
quarks with a tree-level tadpole-improved clover coefficient). They can
see effective-mass plateaus despite using local-local correlation
functions.  They find large dependence on the heavy quark mass.  For
their preliminary calculation they rely on the {\it static\/}
renormalization constants.  It is not clear that the large
mass-dependence will be mitigated upon using the correct heavy-mass
dependent renormalization constants, as was the case for the decay
constant~\cite{GLOK_hl9801038}.

Relying, then, only on conventional results, I quote a world average for
$B_{B}(m_{B})$ in Table~4.  This value has been historically stable; it
agrees with Soni's {\sl Lattice '95\/} estimate~\cite{Son_lat95}.
Indeed, even prehistoric estimates~\cite{Ber88} are good because any
errors in normalization (e.g.\ non-KLM factors) cancel in the ratio.
All groups agree that the light quark mass dependence of $B_{B}$ is
small.
\begin{table}
\begin{center}
\begin{tabular}{lll}
\hline
                                               & Quenched          & ``Unquenched''    \\
\hline					    
\\					    
$B_{B_d}(m_{b})$                               & $0.86(4)(8)$      & $0.86(4)(8)(8)$   \\
$B_{B_s}/B_{B_d}$                              & $1.00(1)(2)$      & $1.00(1)(2)(2)$   \\
$f_{B_d}\sqrt{\hat{B}_{B_d}^{\rm nlo}}$        & $190^{+25}_{-25}$ & $215^{+40}_{-30}$ \\
$\frac{f_{B_s}\sqrt{\hat{B}_{B_s}^{\rm nlo}}}
      {f_{B_d}\sqrt{\hat{B}_{B_d}^{\rm nlo}}}$ & $1.14^{+6}_{-5}$  & $1.14^{+7}_{-6}$  \\
\hline
\end{tabular}
\end{center}
\label{Table:B_average}
\caption{World averages for quenched and ``unquenched'' quantities.
}
\end{table}

Bernard, Blum and Soni (BBS)~\cite{BerBluSon_hl9801039} have recommended
that the $SU(3)$ flavor dependence be investigated by calculating
directly a ratio of matrix elements giving
$\frac{M_{B_{s}}^2 f_{B_s}^2 B_{B_{s}}}
      {M_{B_{s}}^2 f_{B_s}^2 B_{B_{s}}}$
One would expect that many systematic errors would cancel.  Lin and
Lellouch~\cite{Lin_lat98} have since calculated this quantity using
clover quarks (BBS used Wilson).  Both groups find that the direct
method is compatible with, but unfortunately has larger statistical
errors (by a factor of $2$ or $3$) than, the indirect method which
combines separate calculations of $B_{B}$ and $f_{B}$, for both strange
and light quarks.

\section{SEMILEPTONIC DECAYS}

The differential decay rate for $P(Q\bar{q})\longrightarrow
P^{\prime}(q^{\prime}\bar{q}) l\nu_l$ is
\begin{eqnarray*}
\frac{d\Gamma}{dq^2}
& = & 
\frac{G_{F}^2 p^{\prime}{}^3}{24\pi^3}
|V_{q^{\prime}Q}|^2 |f_{+}(q^2)|^2
\end{eqnarray*}
where $f_{+}(q^2)$ is a vector form factor, obtained from the matrix
element
\begin{eqnarray*}
\lefteqn{\langle P^{\prime}(\vec{p}^{\prime}) | V_{\mu} | P(\vec{p}) \rangle}\\
& = & 
      \left( p + p^{\prime} - q\frac{m_P^2-m_{P^{\prime}}^2}{q^2}\right)_\mu 
      f_{+}(q^2)\\
&  & 
      + q_{\mu}\frac{m_P^2-m_{P^{\prime}}^2}{q^2} f_{0}(q^2)
\end{eqnarray*}

Similarly, the differential decay rate for $P(Q\bar{q})\longrightarrow
P^{*}(q^{\prime}\bar{q}) l\nu_l$ is expressed in terms of a vector form
factor $V(q^2)$ and two axial form factors $A_{1}(q^2)$ and
$A_{2}(q^2)$.

Experiments are able to determine an intercept and slope for
$f^{+}|_{(0)}$, so that it is conventional to quote theoretical
predictions at $q^{2}=0$; furthermore, the form factor is often
parameterized with pole dominance form
$f^{+}(q^2)=f^{+}(0)/(1-q^{2}/m^{2})$ to aid interpolation and
extrapolation.  Unfortunately, lattice calculations work best at small
three-momenta, i.e.\ near zero recoil, $q^{2}_{\rm
max}=(m_{P}-m_{P^{\prime}})^2$, and so for decays such as
$B\rightarrow\pi$, large extrapolations are necessary.

Semileptonic form factors were reviewed by T. Onogi at {\sl Lattice
'97}~\cite{Ono_lat97}, by J. Flynn~\cite{Fly_lat96} at {\sl Lattice
'96\/} and by J. Simone~\cite{Sim_lat95} at {\sl Lattice '95}.  For a
very comprehensive and recent survey of lattice results, see the review
by Flynn and Sachrajda~\cite{FlySac_hl9710057} (on which the discussion
of the present section relies heavily).

\subsection{Semileptonic $D\rightarrow K,K^*$ Decays}

For semileptonic $D\rightarrow K,K^*$ decays, the maximum meson recoil
momenta are below $1\,{\rm GeV}/c$ so all $q^{2}$ can be sampled with
small momentum-dependent discretization errors; thus one can {\it
interpolate\/} to $q^2=0$. Furthermore, one can sit on $s$ and $c$ so
the only mass-extrapolation needed is for the light quark.  Thus, $D$
decays are an excellent test for the lattice; experimental
determinations of $|V_{cs}|$ are already good and are likely to improve
soon.  Some disadvantages of the lattice calculations are that the charm
and strange quark masses are non-degenerate, so there is no natural
normalization condition, and that care need be taken to reduce
discretization errors for the charm mass.

\FIGURE
  {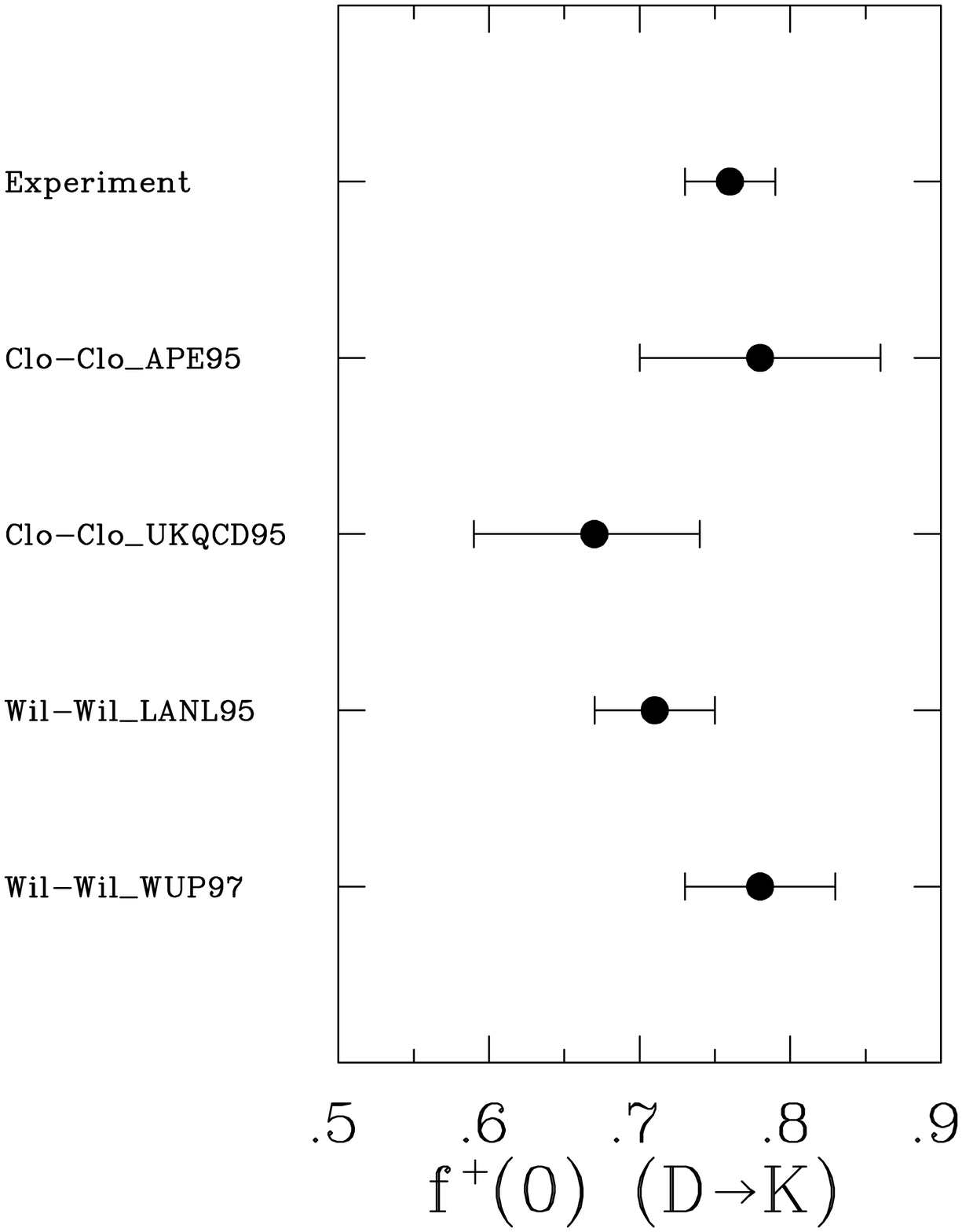}
  {DtoKfplus}
  {Lattice estimates of $f^{+}(0)$ from APE~\cite{APE_hl9411011},
   UKQCD~\cite{UKQCD95}, LANL~\cite{LANL_hl9512007} and
   Wuppertal~\cite{Wup97}, compared to experiment~\cite{DtoKex97}.}
Figure~\ref{Fig:DtoKfplus} compares a recent experimental
estimate~\cite{DtoKex97} with a collection of lattice results for
$f^{+}(0)$, from APE~\cite{APE_hl9411011}, UKQCD~\cite{UKQCD95},
LANL~\cite{LANL_hl9512007} and Wuppertal~\cite{Wup97}.
(See~\cite{FlySac_hl9710057} for a more complete summary including older
results; those listed here use either the SW action, or Wilson with KLM
normalization.)  The agreement is quite satisfactory.  The agreement for
the $D\rightarrow K^*$ form factors, $V(0)$, $A_{1}(0)$, and $A_{2}(0)$,
is also good, although somewhat poorer for $A_1$ and $A_2$, which depend
upon correctly normalizing the lattice axial current.

All of these results are from quenched calculations without a continuum
extrapolation.  A new effort this year seeks to remedy this deficiency.
Preliminary results from FNAL were presented by J. Simone at this
conference~\cite{Sim_lat98}.  Their quenched calculation uses the SW
(clover) action interpreted non-relativistically in the ``Fermilab
approach''; thus they can ``sit on the charm quark''.  At this early
stage in their project, the light quark has strange-quark mass.  They
see a gentle $a$-dependence of the matrix elements which bodes well for
controlling the continuum extrapolation.  Their goal is to precisely
determine decay rates for $D \rightarrow \pi l \nu$ and $D \rightarrow K
l \nu$; the {\sl FOCUS\/} experiment will soon have a high-statistics
determination of BR($D \rightarrow \pi l \nu$)/BR($D
\rightarrow K l \nu$).

Preliminary results from UKQCD were presented by C. Maynard at this
conference~\cite{May_lat98}.  Their quenched calculation on $24\times48$
lattices ($\beta=6.2$) uses a ${\cal O}(a)$-improved SW with a
non-perturbative value of clover coefficient, and simulates with a heavy
quark mass near charm, and a light quark mass both near strange and
close to the chiral limit.  $f^{+}$, $f^{0}$ are simultaneously fit (to
pole-dominance model) for spatial and temporal components of vector
current.  The statistics are quite competitive: $3$--$5\%$ errors for
$D\rightarrow K$ (from $216$ configurations).  The results are to be
compared with older results which used tree-level clover-coefficients.

\subsection{Semileptonic $B\rightarrow \pi,\rho$ Decays}

Semileptonic decays of the $B$ are very important for constraining CKM
matrix elements.  Lattice calculations are really needed since HQET
symmetry is not as helpful as for $B\rightarrow D$.  Unfortunately, the
large bottom mass means that simulations must be at the charm mass and
then extrapolated to bottom, else care must be taken to design an action
for which discretization errors are small and controlled.  Furthermore,
the maximum meson recoil momenta are large so all $q^{2}$ cannot be
sampled (contrast $D$ decays) without large momentum-dependent
discretization errors.  Thus large extrapolation to $q^2=0$ are made for
which one must rely on the pole dominance model.

Although heavy quark symmetry says nothing about overall normalizations, it
does predict that hadronic matrix elements for light-meson decay modes such as
$P\rightarrow \pi l\nu$, scale in the heavy quark limit, $M \rightarrow
\infty$, for $q^2$ near $q^{2}_{\rm max}$ (zero recoil).  Form factor scaling
follows: the leading mass dependence is $f^{0}\sim M^{-1/2}$, $f^{+}\sim
M^{+1/2}$ for $B\rightarrow \pi l \nu$, and $V\sim M^{+1/2}$, $A_1\sim
M^{-1/2}$, and $A_2\sim M^{+1/2}$ and for $B\rightarrow \rho l \nu$.
There is a consensus (see T. Onogi's review~\cite{Ono_lat97} at last
year's lattice conference) that such scaling is seen in lattice
simulations for $q^{2}_{\rm max}$, and thus can be used to extrapolate
simulated form factors to the $B$ mass.

\FIGURE
  {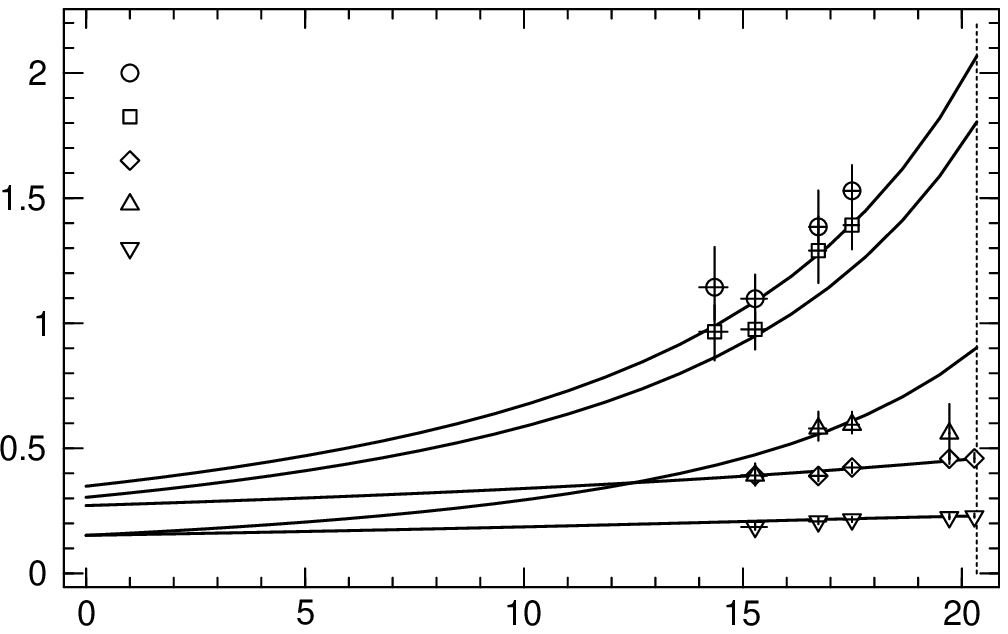}
  {UKQCD_Btorho}
  {Fit to lattice results for $V$, $A_0$, $A_1$, $T_1$, $T_2$, versus
   $(q/{\rm GeV})^2$, for $\rho$ final state.
   From~\cite{UKQCD_hl9708008}.}
To obtain results at smaller $q^2$, one has to extrapolate aided by a
pole model.  UKQCD~\cite{UKQCD_hl9708008} has reanalyzed some older data
and exploited the scaling constraints of the light-cone sum rule to
guide model-dependent extrapolation to $q^2=0$ where the leading mass
dependence for all form factors is $\sim M^{-3/2}$.  (They point out
that the analyses of some other groups violate these scaling relations.)
Figure~\ref{Fig:UKQCD_Btorho} show the results of a simultaneous fit
which respects all constraints.  One can see that the extrapolation to
$q^{2}=0$ is very large indeed.

\FIGURE
  {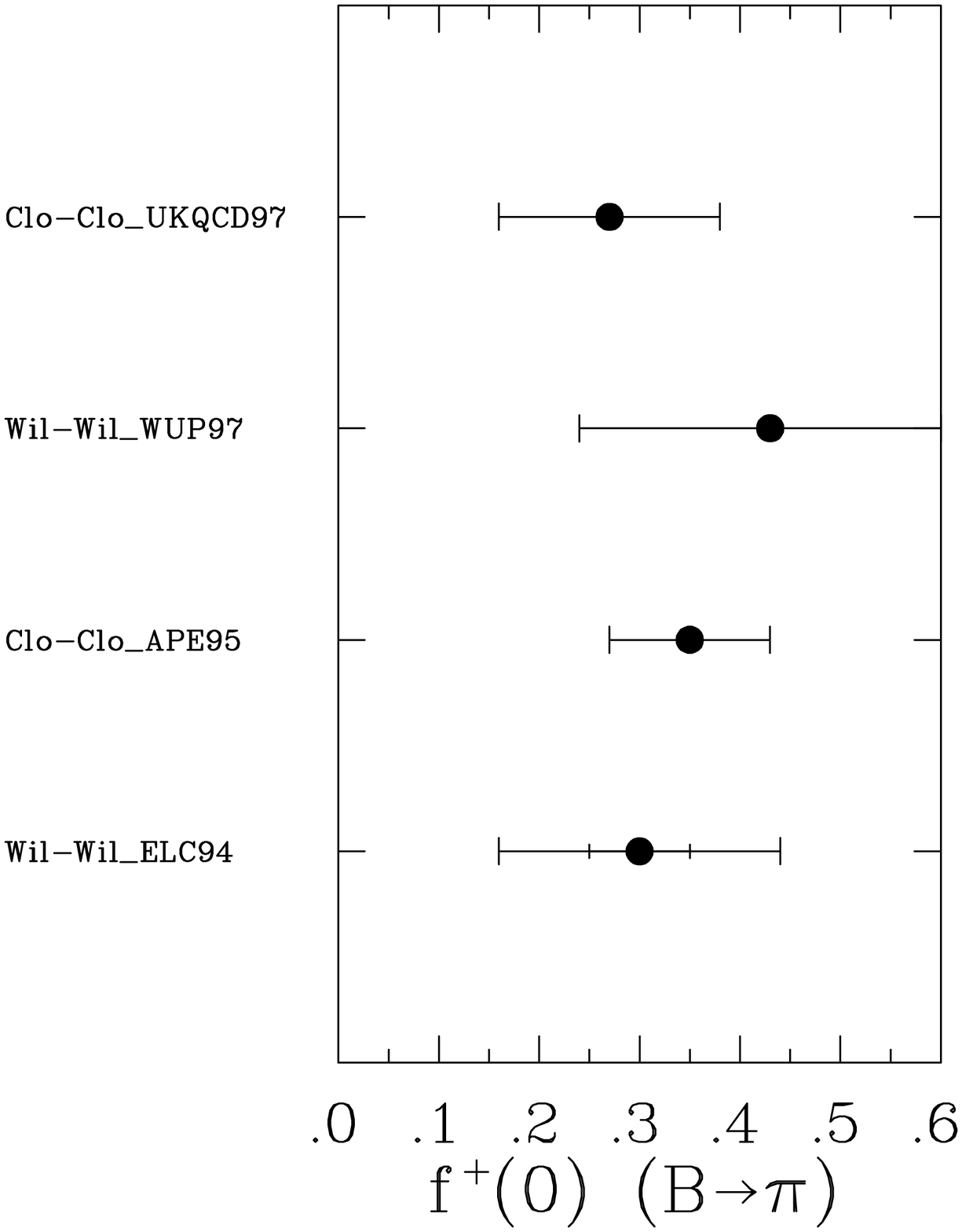}
  {Btopi_f}
  {Lattice estimates of $B\rightarrow \pi$ $f^{+}(0)$ from
   UKQCD~\cite{UKQCD_hl9708008}, Wuppertal~\cite{Wup97},
   APE~\cite{APE_hl9411011} and ELC~\cite{ELC94}.}
Figure~\ref{Fig:Btopi_f} compares previous lattice estimates of $f^{+}$
at $q^2=0$ from UKQCD~\cite{UKQCD_hl9708008}, Wuppertal~\cite{Wup97},
APE~\cite{APE_hl9411011} and ELC~\cite{ELC94}.

\FIGURE
  {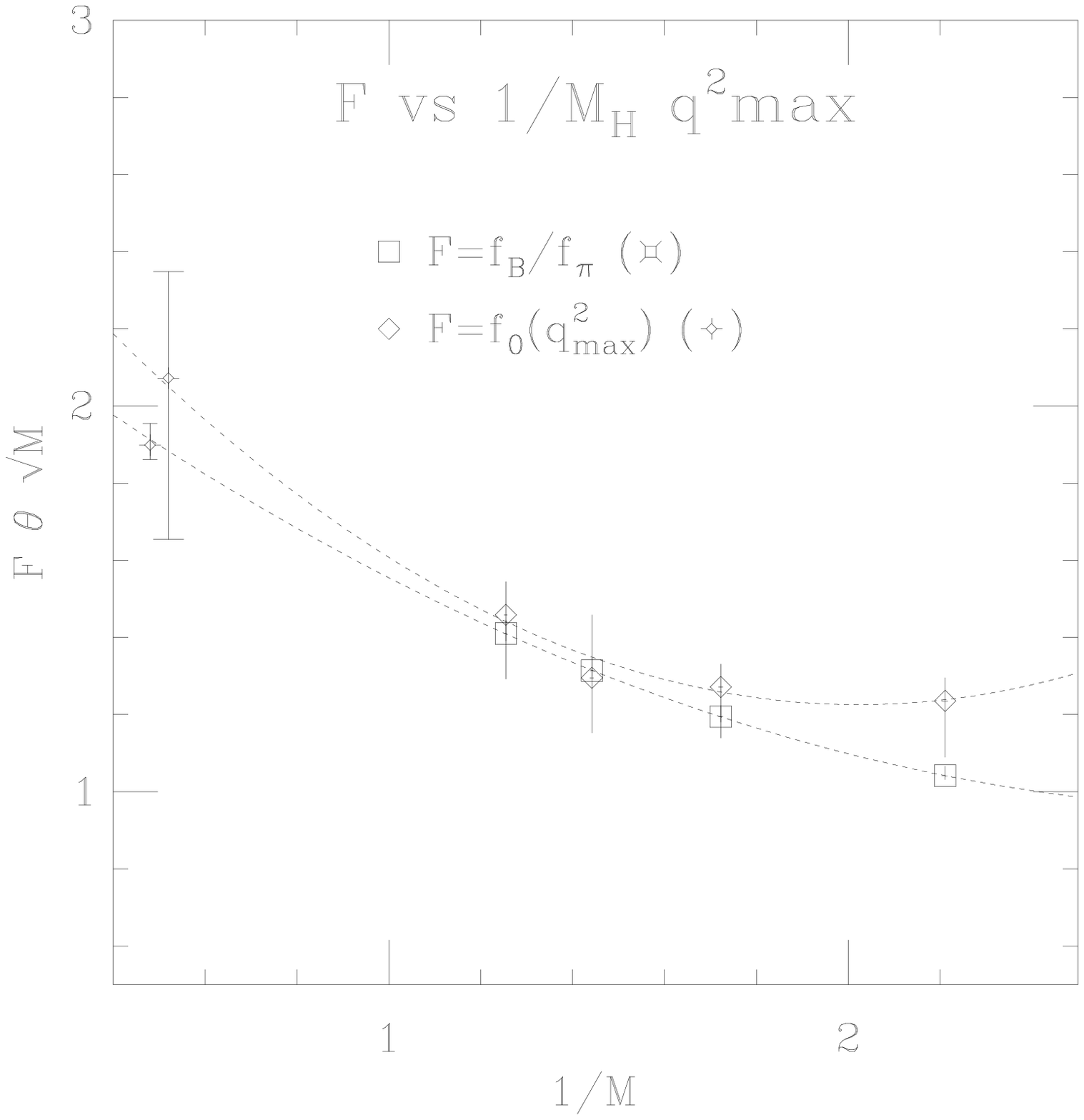}
  {UKQCD_softpion}
  {Comparison of $f^{0}(q^2_{\rm max})$ and $f_{B}/f_{\pi}$.
   From~\cite{May_lat98}.}
Preliminary results from UKQCD were presented by C. Maynard at this
conference~\cite{May_lat98}, in conjunction with their semileptonic $D$
decay calculation (quenched, $24\times48$, $\beta=6.2$, ${\cal
O}(a)$-improved SW with a non-perturbative clover coefficient, heavy
quark mass near charm).  At last year's conference, we saw
evidence~\cite{Ono_lat97} of a large violation of the soft-pion theorem
which predicts the relation
\begin{eqnarray*}
      f^{0}(q^2_{\rm max})
& = & 
      f_{B}/f_{\pi}
\end{eqnarray*}
This year, UKQCD sees no such violation as demonstrated in
Figure~\ref{Fig:UKQCD_softpion}.
The UKQCD calculation includes renormalization constants, and uses a
smaller lattice spacing [$\beta=6.2$, versus $\beta=5.9$ for
JLQCD~\cite{JLQCD_Tom_lat97} (SW action) and $\beta=5.8$ for
Hiroshima~\cite{Mat_lat97} (Wilson light quarks, NRQCD heavy)].

FNAL, as presented by S. Ryan at this conference, are engaged in a
multi-$a$ calculation of $B$ semileptonic decays, in conjunction with
their semileptonic $D$ decay calculation, with the intention of taking
the chiral and continuum limits.  As described in the previous
subsection, the advantage of using their action is that they can sit at
the bottom mass (or any other), and avoid having to extrapolate in the
heavy quark mass.  They verify that the heavy-mass dependence of the
matrix elements is very gentle, from the charm through the bottom
regime.  Of course, they are subject to the same limitations as other
groups for extrapolations in the light quark mass, and in $q^{2}$.

A novel approach~\cite{Agl_hp0806277} has been proposed to predict form
factors for exclusive processes such as $B\rightarrow\pi$ by computing
light-cone wave functions (not, as before, just their moments) in terms
of lattice correlation functions.  Although no simulation results have
been presented yet, the method is intriguing because it can be used at
small $q^{2}$ directly.

In summary, all groups agree on the $1/M$ scaling near $q^{2}_{\rm
max}$.  ELC, APE and UKQCD use scaling to extrapolate from charm to
bottom; JLQCD, Hiroshima and FNAL can sit on the bottom. Estimates of
the form factor at $q^{2}=0$ have a large extrapolation, which is model
dependent.  Continuum extrapolations have not been published, but are in
the works.  A comprehensive estimate of systematic errors, as for $f_B$,
has yet to be made.  One can hope that future estimates of systematic
errors might follow improvements in decay constant calculations, with a
lag of two or three years.  It will continue to be difficult to simulate
at low $q^2$, but it may not be necessary.  Experimental data may soon
be available at high $q^2$; if so, a comparison with accurate lattice
results will be sufficient to determine $|V_{ub}|$ with impressive
precision.

\subsection{Semileptonic $B\rightarrow D,D^*$ Decays}

Semileptonic decays $B\rightarrow D,D^*$ are used to extract the
$|V_{cb}|$ CKM matrix element.  Heavy quark symmetry plays a central
role.  The intuitive picture is well known. (See
reference~\cite{Neu_hp9306320}, for example).  Typical momenta exchanged
between heavy and light quarks are ${\cal O}(\Lambda_{\rm QCD})$; light
degrees of freedom cannot resolve distances $1/m_Q \ll 1/\Lambda_{QCD}$
and are blind to the mass and spin of heavy quark; thus hadronic systems
which differ only in the mass and spin of the heavy quark have the same
light {\it dof\/} configuration.  The non-perturbative light {\it dof\/}
(``brown muck'') interacts with heavy-quark static color field (which
extends over long distances).  When an external weak current boosts the
heavy quark ($m_b\rightarrow m_c$), the brown muck reacts and the form
factor suppression is {\sl universal}.

Heavy quark symmetry suggests rewriting the matrix elements in terms of
form factors which are functions of velocity transfer $\omega=v\cdot
v^{'}$, rather than momentum transfer.  Thus $\langle P^{\prime} \ |
V_{\mu} \ | \ P \rangle
\propto (v+v^{\prime}) h_{+}(\omega) + a(v-v^{\prime})
h_{-}(\omega)$. Similarly, $\langle P^{*} | V_{\mu} | P \rangle$ is expressed
in terms of $h_{V}$ and $\langle P^{*} | A_{\mu} | P \rangle$ in terms of
$h_{A_{1}}$, $h_{A_{2}}$, $h_{A_{3}}$. Then
\begin{eqnarray*}
      h_{i}(\omega) 
& = & 
      \left[ \alpha_i + \beta_i(m_b,m_c,\omega) 
                      + {\cal O}(\Lambda/m_Q)\right]\xi(\omega)
\end{eqnarray*}
where the HQS symmetry enforces $\alpha_{+,V,A_1,A_3}=1$ and
$\alpha_{-,A_2}=0$~\cite{Neu_hp9306320}.  $\xi(\omega)$ is the famous
Isgur-Wise function.

The experimental measurement of the differential cross sections yields
$|V_{cb}|^2 {\cal F}(\omega)^2$ where the physical form factor, ${\cal
F}(\omega)$, equals $\xi(\omega)$ plus perturbative and power
corrections.  Experimentally, the shape of ${\cal F}(\omega)$ is not
well known, and so it is common to parameterize the extrapolation of the
data back to zero recoil (where a theoretical calculation of the
normalization ${\cal F}(1)$ would then determine $|V_{cb}|$), in terms
of the slope at zero recoil, $-\hat{\rho}^2$.  A recent world
(experimental) average is $\hat{\rho}^{2} =
0.75(11)$~\cite{Gib_he9704017}.  A more precise theoretically-determined
value would greatly aid the extrapolation back to zero recoil.  A task
for lattice calculations is then a determination of $\hat{\rho}^2$, or
of the corresponding slope, $-\rho^2$, of the Isgur-Wise function to
which it is related by calculable corrections~\cite{Neu_hp9306320}.

The first lattice calculations of the Isgur-Wise function were reviewed
by Kenway at {\sl Lattice '93}.  (See also the aforementioned reviews by
J. Simone~\cite{Sim_lat95} and by Flynn \&
Sachrajda~\cite{FlySac_hl9710057}.)  Since then, there have been
surprisingly few lattice calculations, some of which are listed in
Table~5.  There also exist
calculations~\cite{ManOlg_hl9312013,DraMcN_hl9509060,HasMat_hl9511027}
which simulate the heavy quark effective theory directly on the lattice,
but renormalization is subtle, and discretization errors can be sizeable
for lowest-order actions on modest-sized lattices.
\begin{table}
{\setlength{\tabcolsep}{3.5pt}
\begin{tabular}{llll}
\hline
         & $\rho^{2}_{d}$            & $\rho^{2}_{s}$            & $|V_{cb}|\times 10^2$ \\
\hline
BSS~\cite{BerSheSon93}
         &                           & $1.4(2)(4)$               & $4.4(5)(7)$           \\
UKQCD~\cite{UKQCD95b}
         & $0.9(^{2}_{3})(^{4}_{2})$ & $1.2(^{2}_{2})(^{2}_{1})$ & $3.7(1)(2)(^{4}_{1})$ \\
LANL~\cite{LANL_hl9512007}
         & $0.97(6)$                 &                           &                       \\
\hline
\end{tabular}
}
\caption{A collection of lattice results for the slope of the Isgur-Wise
function. This is then used to constrain extrapolation of the
experimental data to obtain the CKM matrix element listed.}
\label{Table:Isgur-Wise}
\end{table}

There has been renewed interest recently, and at this conference we had
reports of simulations from FNAL (presented by
S. Hashimoto~\cite{Has_lat98}) and UKQCD (D. Richards~\cite{Ric_lat98}).
A. Kronfeld~\cite{Kro_lat98}, P. Boyle~\cite{Boy_lat98} and
J. Sloan~\cite{Slo_lat98} reported on renormalization calculations.

The UKQCD calculation is part of a larger calculation including
$D\rightarrow K, K^*$ and $B\rightarrow \pi,\rho$ form factors as
described above.  Recall that the clover coefficient is determined
non-perturbatively.  The matching coefficient in the renormalized vector
current $V^{\rm R}_{\mu} = Z_V ( 1 + b_V a m_q) [ V_{\mu}^{\rm latt} +
\cdots]$ can be computed for degenerative transitions at zero momentum,
and then compared with the ALPHA collaboration's~\cite{ALPHA_hl9605038}
non-perturbative determination of $Z_V$ and $b_V$.  The striking
agreement leads UKQCD to believe that they have good understanding of
the discretization errors.

S. Hashimoto~\cite{Has_lat98} presented preliminary FNAL results for
$B\rightarrow D$, using the same action as for the other semileptonic
form factor calculations previously described.  Recall they can simulate
at the $b$ mass; currently, the light quark mass is around strange.
They consider a {\sl ratio\/} of matrix elements at zero recoil
\begin{eqnarray*}
      |h_{+}(1)^{B\rightarrow D}|^2
& = & \frac{
        \langle D | \bar{c}\gamma^0 b | B \rangle 
        \langle B | \bar{b}\gamma^0 c | D \rangle} 
           {
        \langle D | \bar{c}\gamma^0 c | D \rangle 
        \langle B | \bar{b}\gamma^0 b | B \rangle}
\end{eqnarray*}
for which statistical errors are small, current renormalization and
systematic errors largely cancel.  (This is the same ratio as first
defined by Mandula and Ogilvie~\cite{ManOlg_hl9312013} and then used by
Draper and McNeile~\cite{DraMcN_hl9509060} for lattice HQET.)  Using a
corresponding ratio to obtain $h_{-}$, they then construct a linear
combination to obtain ${\cal F}(1)$, i.e.\ the physical form factor at
zero recoil!  Previously, lattice calculations had predicted only the
slope to constrain the fit to experimental data leaving ${\cal
F}(1)|V_{cb}|$ as the free parameter, and then relied on sum rule
calculations of ${\cal F}(1)$.  With this development, HQET and the
lattice alone can determine $|V_{cb}|$.

\section{HIGHLIGHTS FROM QUARKONIA}

At this conference we saw several interesting calculations of quarkonia.
I will not attempt to put these results in context by making a
comparison with a complete list of earlier results, due to space
constraints, but rather will take this opportunity to advertise several
of these calculations.

Presently, the usual approaches used in calculating properties of
quarkonia on the lattice are NRQCD (expansion in $v^2$; breaks down for
small masses and small lattice spacing) and the Fermilab approach
(improve at $am\gg 1$; coefficients are mass-dependent; present
simulations with ``non-relativistic reinterpretation'' of Wilson-SW
action).

Problems show up in charmonium hyperfine splitting.  At this conference,
Shakespeare and Trottier~\cite{ShaTro_hl9802038} compared the effect of
${\cal O}(v^6)$ versus ${\cal O}(v^4)$ terms, and the effect of the
detailed way in which mean field is invoked (mean-link from Landau gauge
versus from fourth-root-of-plaquette tadpole improvement).  They see
NRQCD breakdown for $aM_c^0<1$ (as expected) and see a clear preference
for using the Landau prescription.  It should be noted that this is not
an indictment of NRQCD, nor of mean-field improvement.  Calculations of
bottomonium are well under control.  For charmonium, the effect of
higher order terms is as expected, except that the prefactors are a
factor of two or so higher than one would predict from dimensional
analysis.

T. Klassen~\cite{Kla_lat98} showed us the first results from a long-term
program which provides an alternative to NRQCD or Fermilab: Symanzik
improvement on anisotropic $\xi=a_s/a_t \gg 1$ lattices.  His quark
action shares some formal similarities with that of Fermilab.  It has
on-shell ${\cal O}(a)$ classically-improvement plus a non-perturbative
${\cal O}(a)$ improvement which requires a tuning of the Wilson
parameter $r(m,\xi)$.  However, the coefficients of the electric and
magnetic terms in the action are mass-independent (contrast Fermilab).
For his first calculation, he uses mean-field (Landau) to estimate these
coefficients, and obtains $r(m,\xi)$ from charmonium dispersion.  His
continuum extrapolation is well-behaved.  He sees spectacular agreement
with experiment for the charmonium $P$-state splitting.

R. Horgan~\cite{Horg_lat98} presented preliminary results of a long-term
project using an anisotropic lattice for NRQCD\@.  Their goal is a high
statistics calculation for bottomonium and for hybrid mesons.  Early
results are encouraging and are in good agreement with those from
isotropic lattices.

K. Hornbostel~\cite{Horn_lat98} presented a new analysis from the NRQCD
collaboration of the bottom (and charm) mass.  They used a
tadpole-improved NRQCD action including next-to-leading order
relativistic and discretization corrections, computed spin-averaged
$1P$--$1S$ splittings to determine the scale $a$, tuned the bare mass so
the kinetic mass from the dispersion relation agreed with experiment,
perturbatively related the bare mass to the $\overline{MS}$ mass (using
a Lepage-Mackenzie $q^*$), and ran the $\overline{MS}$ mass to its own
scale.  They obtain $4.28(3)(3)\,{\rm GeV}$ from their $\beta=6.0$ data
(which has the best statistics).

Bali and Boyle~\cite{BalBoy_hl9809180} have done some nice work in which
they measure the potential on the lattice, and then parameterize it to
solve the Schr\"{o}dinger equation on (in)finite lattices. They then
vary the lattice volume and quark masses to deduce from their model
qualitative features seen in the full simulations; namely,
non-negligible finite-size effects, a non-trivial dependence on quark
mass, and splittings strongly dependent on the Coulomb coupling.

I would also like to advertise a calculation of the heavy-quark leptonic
width using NRQCD by Jones and Woloshyn~\cite{Jon_lat98}, a new estimate
of $\alpha_s$ from the SESAM collaboration~\cite{Spi_lat98}, and a study
by Fingberg~\cite{Fin_lat98} of the effect of the gluon condensate on
the low-lying quarkonium spectrum.  Pennanen~\cite{Pen_lat98} presented
results from a study of four-quark systems, while
Morningstar~\cite{Mor_lat98} presented final results for the presence of
gluonic excitations in the presence of a static quark-antiquark pair.

\section{CONCLUSIONS}

The use of improved actions, techniques, analysis and computer power in
recent years has been most fruitful.  Lattice QCD has made very useful
and competitive predictions of many phenomenological quantities crucial
for the confrontation of theory and experiment.

It has been a banner year for the calculation of heavy-light leptonic
decay constants such as $f_{B}$.  There are now reliable estimates of
systematic errors for the quenched calculation, and there is good
agreement among groups using a variety of methods.  In the near future,
we should see if the final analysis (including a continuum
extrapolation) from conventionally-interpreted SW actions (which use a
non-perturbative estimate of the clover coefficient) gives results which
concur with the clustering of several results obtained from
non-relativistic interpretation.  If so, attention must turn to the
remaining formidable issue of more reliably estimating the effects of
the quenched approximation.  With new actions and methods, more
attention should be paid to the heavy-light spectrum, where current
estimates of the sensitive hyperfine splitting are too small.

Estimates of the $B_{B}$ parameter from neutral $B$ mixing seem rather
robust; since it is determined from a ratio of correlation functions,
systematic errors and statistical fluctuations largely cancel.  More
modern estimates using Wilson or SW actions agree with older results.
The vacuum saturation, or factorization, approximation appears to hold
at the $15\%$ level.  Earlier variance among results using the static
approximation for the heavy quark are better understood, and somewhat
reduced.  Simulations with an NRQCD heavy quark were presented this
year, but need a difficult renormalization calculation for completeness.

For semileptonic decays, we need need more complete estimates of systematic
errors (continuum limit, variety of actions).  It would be nice to see more
groups calculating these form factors.  The kinematics of $D \rightarrow K,K^*$
provide a good place to test actions, techniques and analysis.  For $B
\rightarrow \pi, \rho$, the kinematics demand a large, model-dependent,
extrapolation in $q^2$.  (One must also beware discretization errors due
to the large bottom mass; the two philosophies used are the same as for
leptonic decay calculations: either attempt to control errors with a
non-relativistic interpretation of the action, or simulate around the
charm mass, and boldly extrapolate.)  It is unlikely that the large
$q^2$ extrapolation can be avoided in the near future with conventional
approaches. It may not be necessary.  We expect our experimental
colleagues, using increased luminosities available around the turn of
the millennium, to be successful in measuring the form factors at high
$q^2$.  Then a direct comparison of theory and lattice results should
provide a good estimate of $|V_{ub}|$.

The power of heavy-quark-symmetry applied to $B \rightarrow D$
semileptonic decays has reduced the phenomenologist's dependence on
lattice estimates, but the lattice can still make fruitful contributions
by providing a reliable estimate of the slope and intercept of the form
factor at zero recoil.  New calculations using relativistic heavy quarks
have been presented at this conference; it is welcome to see renewed
interest in these form factors.  We expect NRQCD estimates to be
available soon and anticipate that within a year or two these,
relativistic estimates, and estimates from directly simulating HQET,
will all be reconciled as has been the case for leptonic decay
constants.

Calculations of quarkonia continue to be a good forum for testing new
actions and techniques and for predicting parameters of the standard
model; we saw updated estimates of $\alpha_s$ and $m_b$ at this
conference.  Bottomonium calculations are in good shape; NRQCD
calculations in particular are quite precise.  The convergence of the
expansion in inverse mass is poor for charmomium, however, and so this
is a good place to try new improved actions.  We saw that anisotropic
actions look very promising.

Calculations of heavy quark physics on the lattice have reached a new
level of maturity in the last few years, and have provided the wider
community with reliable non-perturbative estimates of several quantities
of phenomenological interest.  But there are still many quantities for
which it is difficult to be as precise.  Meanwhile, new approaches and
techniques mean that the lattice calculations will continue to
interesting and important for years to come.  Onward and upward!

\section{ACKNOWLEDGEMENTS}

I thank 
A. Ali Khan, 
G. Bali,
C. Bernard, 
P. Boyle,
J. Christensen,
M. Di Pierro,
S. Hashimoto,
J. Hein,
R. Horgan,
K. Hornbostel,
K.-I. Ishikawa,
T. Klassen,
A. Kronfeld,
R. Lewis, 
K.F. Liu,
C.-J. Lin,
G. Martinelli,
C. Maynard,
C. McNeile,
P. Pennanen,
C. Sachrajda,
N. Shakespeare,
H. Shanahan,
J. Shigemitsu,
J. Simone,
D. Sinclair,
J. Sloan,
R. Sommer,
D. Richards,
J. Rosner,
S. Ryan,
A. Ukawa,
R. Woloshyn,
and
N. Yamada
for helpful comments and suggestions and for communicating to me results
of their work prior to publication.

This work is supported in part by the U.S. Department of Energy under
grant numbers DE-FG05-84ER40154 and DE-FC02-91ER75661 and by the Center
for Computational Sciences, University of Kentucky.

\end{document}